\documentclass[fleqn,12pt,twoside]{article}
\usepackage{espcrc1}

\usepackage{graphicx}
\usepackage[figuresright]{rotating}

\newcommand{\AmS}{{\protect\the\textfont2
  A\kern-.1667em\lower.5ex\hbox{M}\kern-.125emS}}

\def\si{{}^1\kern-.14em S_0}
\def\siii{{}^3\kern-.14em S_1}
\def\diii{{}^3\kern-.14em D_1}

\title{Nuclear Physics from QCD}

\author{S.R.~Beane \address{Institute for Nuclear Theory,\\
        University of Washington, Seattle, Washington 98195-1550, USA.}}
       
\begin{document}

% typeset front matter
\maketitle

\begin{abstract}
Recent attempts to make direct contact between QCD and simple nuclear systems
are reviewed.
\end{abstract}

\section{Introduction}

\noindent 
It is now accepted that QCD is the theory which underlies all of
nuclear structure. A fundamental question we may then ask is: How do
nuclear energy levels change as we vary the quark masses in the QCD
Lagrangian?  This is not solely an intellectual exercise; there are
hints that the fundamental parameters of the standard model, such as
$\alpha_{\rm em}$ and $m_q$, may be time-dependent~\cite{Webb}. The
successful predictions of Big-Bang nucleosynthesis (BBN) can be used
to constrain the time dependence of these parameters and thereby
search for physics beyond the standard
model~\cite{yoo}\cite{gail}. However in order to do so one must know
how nuclear physics depends on the fundamental parameters. A more
practical motivation for understanding the quark mass dependence of
nuclear physics, and more generally of hadronic physics, is that in
the near future it is through lattice QCD simulations that definitive
predictions in nuclear physics will be made directly from QCD. And
because lattice QCD is currently simulated with unphysically large
quark masses, there is an inevitable extrapolation to physical quark
masses. If rigorous predictions are to be had from the lattice, this
extrapolation must be controlled in a precise way. Fortunately, the
quark-mass dependence of few-nucleon systems can be studied by
exploiting hierarchies of scales: i.e. by using effective field theory
(EFT)~\cite{BBSvK}\cite{BSmq}\cite{EMmq}.  Given the current state of
technology, simulations of multi-nucleon systems are intractable, but
realistic simulations of two-nucleon systems are feasible. Arguably,
the most promising method is to calculate scattering phase shifts
directly using L\"uscher's finite-volume algorithm, which, for
instance, expresses the ground-state energy of a two-particle state as
a perturbative expansion in the scattering length divided by the size
of the box~\cite{Luscher}. This method has been used successfully to
study $\pi\pi$ scattering~\cite{Gupta:1993rn}. Attempts have been made
to compute nucleon-nucleon (NN) scattering parameters in lattice QCD; Ref.~\cite{fuku}
computes the $\si$ and $\siii$ scattering lengths in quenched QCD
(QQCD) using L\"uscher's method. Of course in the NN system, the
scattering lengths are much larger than the characteristic physical
length scale given by $\Lambda_{\scriptstyle QCD}^{-1}$. Nevertheless
one may expect that at some unphysical value of the pion mass used in
a lattice simulation, the scattering lengths relax to natural values,
thus allowing their determination from the lattice. This is yet
another motivation for understanding the quark-mass dependence of
few-nucleon systems.

Presently, unquenched lattice simulations with the physical values of
the light-quark masses are prohibitively time-consuming, even on the
fastest machines.  While some quenched calculations can be performed
with the physical quark masses there is no limit in which they
reproduce QCD, and consequently they should be considered to be a
warm-up exercise for the ``real thing''.  Relatively recently it was
realized that partially-quenched (PQ) simulations, in which the sea
quarks are more massive than the valence quarks, provide a rigorous
way to determine QCD observables and are less time-consuming than
their QCD counterparts.  It is with PQ simulations that
nuclear-physics observables will first be calculated rigorously from
QCD.  As PQQCD reproduces QCD only in the limit in which the sea-quark
masses become equal to the valence-quark masses which, in turn, are
set equal to the physical values of the light quark masses (we call
this the QCD limit), there are some interesting features of the PQ
theory that are distinct from nature away from the QCD limit. In QCD,
the long-distance component of the NN potential is due to OPE, as
discussed above.  However, in PQQCD there is also a contribution from
the exchange of the $\eta$-meson (in the theory with two flavors of
light quarks). In QCD such an exchange is suppressed due to the large
mass of the $\eta$ compared to the $\pi$. However, in PQQCD the $\eta$
propagator has a double-pole component that depends on the pion mass
due to the hairpin interactions with a coefficient that depends upon
the difference between the masses of the sea and valence quarks.
Therefore, away from the QCD limit the long-distance component of the
NN potential is dominated by one-eta exchange (OEE) and falls
exponentially with a range determined by the pion mass, $\sim
e^{-m_\pi r}$, as opposed to the familiar Yukawa type
behavior~\cite{BSpot}. All is not lost however: it is straightforward to develop the
partially-quenched EFT which matches to a partially-quenched lattice
simulation~\cite{BSnn}.

A second approach to the NN system on the lattice is to study the
simplified problem of two interacting heavy-light
particles~\cite{Richards:2000ix}.
It has been suggested that lattice QCD simulations of the potential
between hadrons containing a heavy quark will provide insight into the
nature of the intermediate-range force between two
nucleons~\cite{JLabMIT}. While the NN potential is not itself an
observable, one may instead consider heavy systems.  In the
heavy-quark limit, the kinetic energy of the heavy hadrons is absent
and the lowest-lying energy eigenvalues, which can be measured on the
lattice, are given by the interaction potential. All discussions to
date have addressed two heavy mesons. This case is somewhat
complicated by the fact that there are degeneracies in the heavy-quark
limit which require a coupled-channel analysis.  By contrast, the
$\Lambda_Q\Lambda_Q$ interaction (where the $\Lambda_Q$ is the
lowest-lying baryon containing one heavy quark, $Q$), does not suffer
from this complication~\cite{ABSl}. Moreover, since the $\Lambda_Q$ is an
isosinglet, there is no OPE, and the leading large-distance behavior
is governed by two-pion exchange (TPE), which is physics analogous to the
intermediate-range attraction in the NN potential. 

In this proceedings I will review recent work in developing the EFTs
relevant to understanding the quark-mass dependence of the NN system
and the $\Lambda_Q\Lambda_Q$ potential, both in QCD and in
PQQCD.

%%%%%%%%%%%%%%
\section{The Quark Mass Dependence of NN}

During the last decade significant effort has been expended in constructing an EFT
to describe nuclear physics.  While it is straightforward to write down all
possible terms in the effective Lagrangian for two or more nucleons,
arriving at a consistent power-counting has proved to be a difficult task.
Weinberg's (W) original proposal~\cite{We90} for an EFT describing
multi-nucleon systems was to determine the NN potentials
using the organizational principles of the well-established EFTs describing
the meson-sector and single-nucleon sector (chiral perturbation theory,
$\chi$PT), and then to insert these potentials into the Schr\"odinger equation
to solve for NN wavefunctions.  Observables are computed as matrix elements of
operators between these wavefunctions.  W power-counting has been extensively
and successfully developed during the past decade to study processes in the
few-nucleon systems~\cite{evgenifb17}.  This method is
intrinsically numerical and is similar in spirit to traditional nuclear-physics
potential theory.  Unfortunately, there are formal inconsistencies in W
power-counting~\cite{KSWa}, in particular, divergences that arise at leading
order (LO) in the chiral expansion cannot be absorbed by the LO operators.
Problems persist at all orders in the chiral expansion, and the correspondence
between divergences and counterterms appears to be lost, leading to
uncontrolled errors in the predictions for observables.  This formal issue was
partially resolved by Kaplan, Savage and Wise (KSW) who introduced a power-counting in
which pions are treated perturbatively~\cite{KSW98}. The NN phase-shifts and
mixing angle in the $\si$ and $\siii-\diii$ coupled-channels have been computed
to next-to-next-to-leading order (N$^2$LO) in the KSW expansion by Fleming,
Mehen and Stewart (FMS)~\cite{FMS} from which it can be concluded that the KSW
expansion converges slowly in the $\si$ channel and does not converge in the
$\siii-\diii$ coupled-channels. Therefore, neither W or KSW power-counting 
provide a complete description of nuclear interactions.
\begin{figure}[!htb]
\begin{center}
\includegraphics*[width=0.4\linewidth,clip=true]{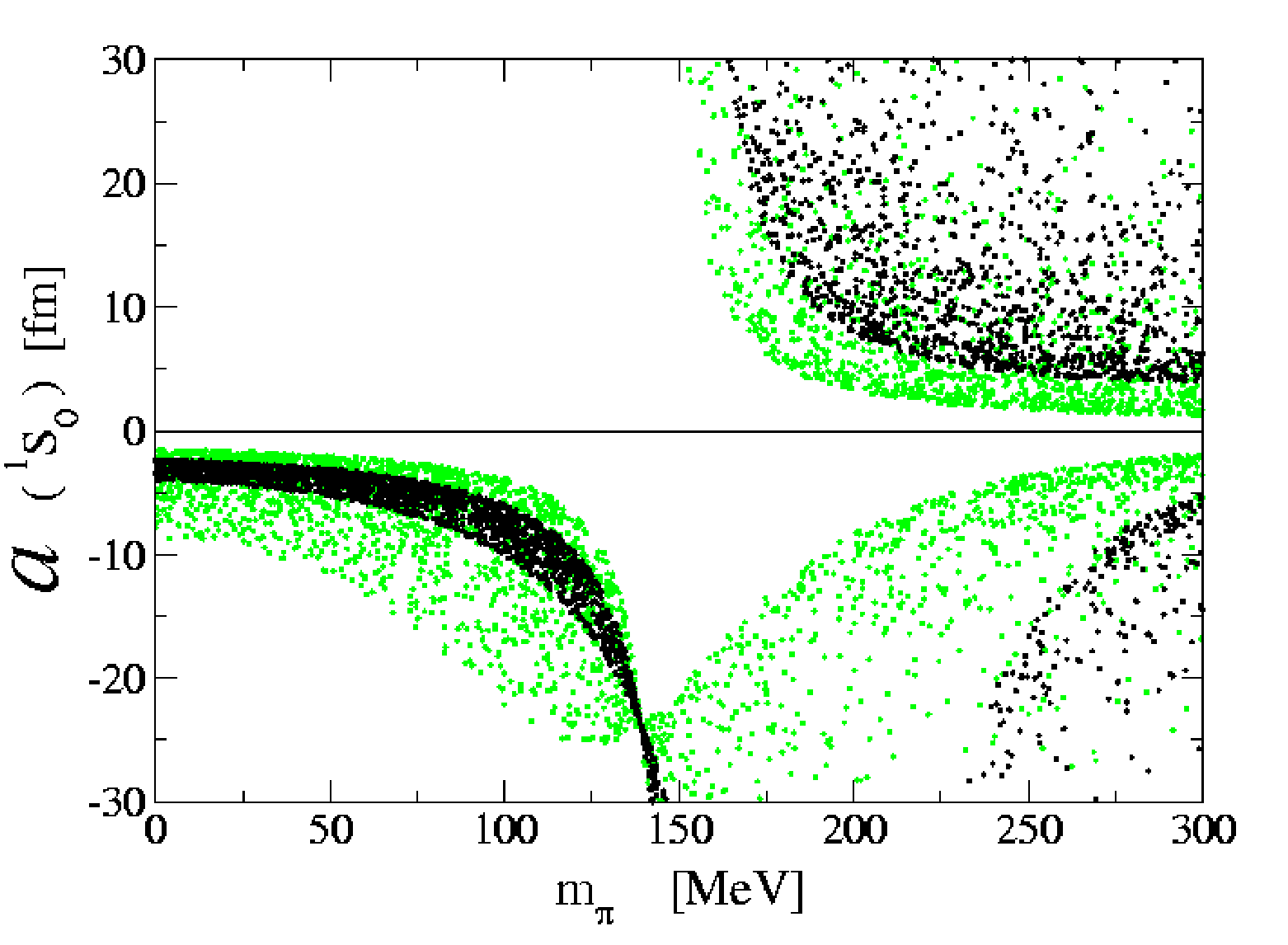}
\hskip 0.6in
\includegraphics*[width=0.4\linewidth,clip=true]{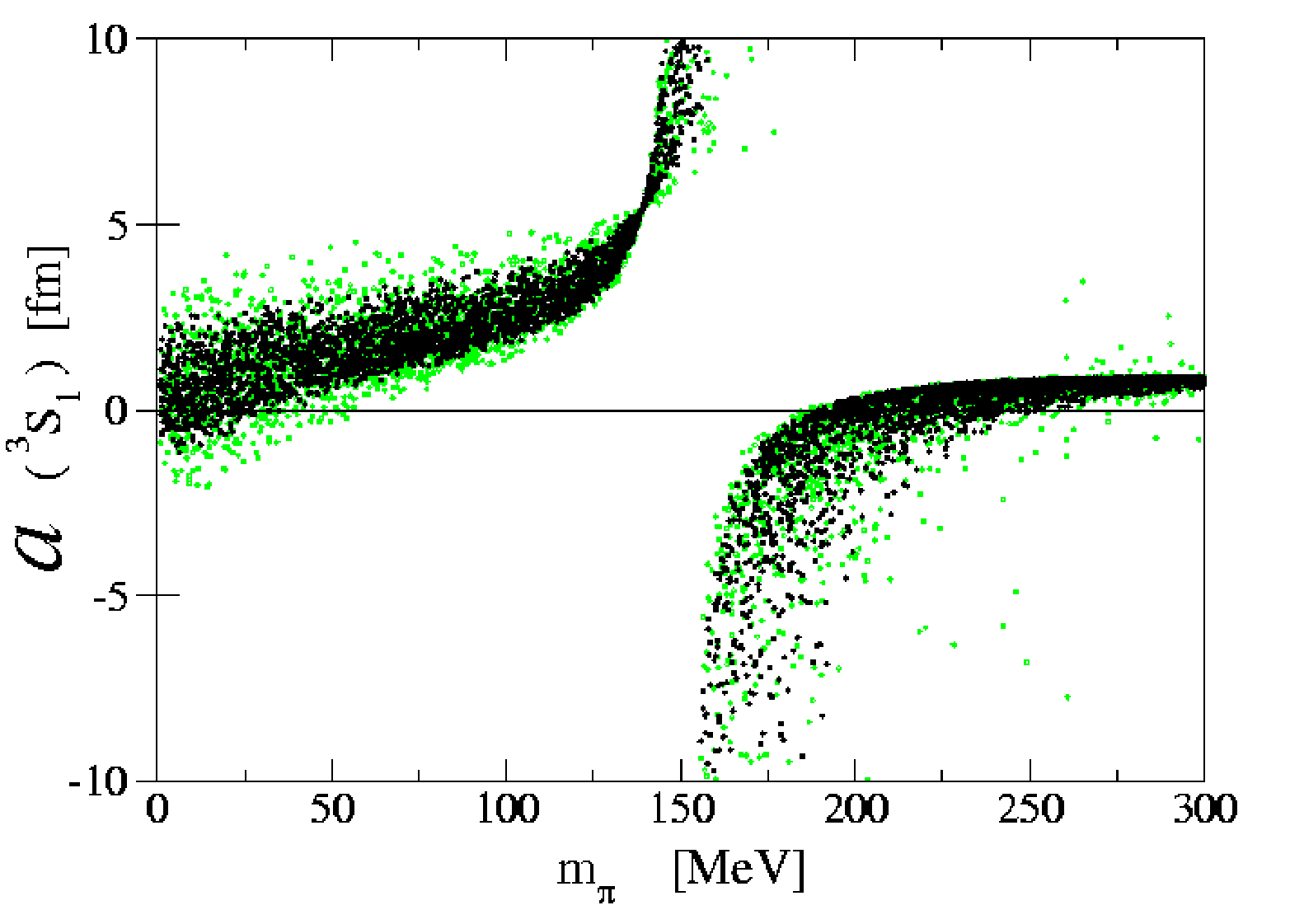}
\end{center}
\vskip -0.5in
\caption{
The left (right) panel shows
the scattering length in the $\si$-channel ($\siii$-channel)
as a function of the  pion mass.
The light gray region corresponds to $\eta=1/5$ and the 
black region corresponds to $\eta=1/15$.
In the $\siii$-channel
the parameter $\overline{d}_{16}$ is taken to be in the interval
$-2.61~{\rm GeV}^{-2} < \overline{d}_{16} < -0.17 ~{\rm GeV}^{-2}$
and $\overline{d}_{18}=-0.51~{\rm GeV}^{-2}$.
 }
\label{fig:1s0mq}
\vskip -0.1in
\end{figure}

The problems with W and KSW power-counting appear to have been
resolved in Ref.~\cite{BBSvK}, which from this point on I will refer to as
BBSvK.  It was realized in FMS that the contributions to the amplitude
that lead to non-convergence in the $\siii-\diii$ coupled-channels
persist in the chiral limit (it is the chiral limit of iterated
one-pion-exchange (OPE) that is troublesome).  Therefore, in BBSvK
power-counting the scattering amplitude is an expansion about the
chiral limit.  This recovers KSW power-counting in the $\si$ channel,
where FMS found it to be slowly converging.  However, in the
$\siii-\diii$ coupled-channels, the chiral limit has contributions
from both local four-nucleon operators and from the chiral limit of
OPE.  It is these two contributions that must be resummed using the
Schr\"odinger equation to provide the LO scattering amplitude in the
$\siii-\diii$ coupled-channels.

In recent papers by Savage and the author~\cite{BSmq} and also by
Epelbaum, Gl\"ockle and Mei\ss ner~\cite{EMmq} EFT was used to determine
the $m_q$-dependence of scattering in the two-nucleon sector. Remarkably, in the
$\si$-channel KSW power-counting can be used to derive an analytic
expression for the scattering length,
\begin{eqnarray}
\frac{1}{a^{(\si)}}=\gamma +\frac{g_A^2 M_N}{8\pi f_\pi^2}\left[ m_\pi^2\,
  \log{\left({\mu\over m_\pi}\right)}+ (\gamma - m_\pi )^2
-(\gamma - \mu )^2 \right] 
-\frac{M_N m_\pi^2}{4\pi}(\gamma -\mu )^2 \ D_2  
\ \ ,
\end{eqnarray}
\begin{figure}[!htb]
\begin{center}
\includegraphics*[angle=-90,width=0.5\linewidth,clip=true]{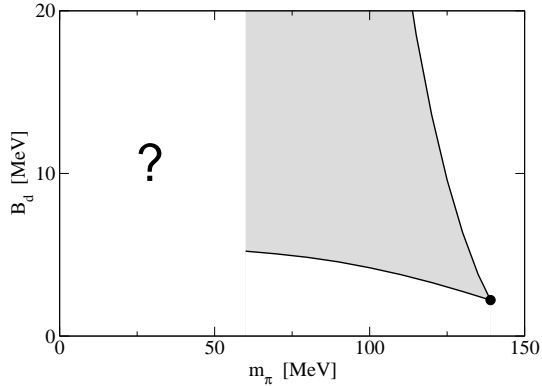}
\end{center}
\vskip -0.5in
\caption{The deuteron binding energy as a function of the 
pion mass (from the physical value to the chiral limit). 
We do not show values below $m_\pi =60~{\rm MeV}$
since the deuteron can be both bound and unbound.
The shaded region corresponds to $\eta=1/5$, 
$-2.61~{\rm GeV}^{-2} < \overline{d}_{16} 
< -0.17 ~{\rm GeV}^{-2}$, and 
$-1.54~{\rm GeV}^{-2} < \overline{d}_{18} < -0.51 ~{\rm GeV}^{-2}$.
 }
\label{fig:deutbe}
\vskip -0.1in
\end{figure}
where $\gamma$ is a LO constant and $D_2 (\mu)$ is a combination 
of  coefficients of operators with a single insertion of $m_q$.
Unfortunately, $D_2$ is presently unknown (in the $\siii-\diii$ channel
$D_2$ contributes to, for instance, $\pi$-deuteron scattering, however
only at one order beyond the current state-of-the-art~\cite{Beane:2002wq}\cite{ulffb17}).
The best that one can do at present is to use naive dimensional
analysis (NDA) to estimate a range of reasonable values for $D_2$,
defined by a parameter $\eta \ll 1$~\cite{BSmq}.
The results of NDA are shown in Fig.~\ref{fig:1s0mq}.
We use scatter plots as the point density represents the probability associated
with a particular set of low-energy constants.
NDA suggests that the di-neutron remains unbound in the 
chiral limit, while a relatively small increase in $m_q$ could lead to a 
bound di-neutron.

In the $\siii-\diii$ coupled channels the situation is somewhat more
complicated.  At NLO in BBSvK counting there is not only OPE,
but also the chiral limit of TPE.  As a consequence,
there are additional counterterms in the single nucleon sector that
contribute in this channel but do not contribute in the $\si$ channel,
in particular $\overline{d}_{18}$ and $\overline{d}_{16}$ associated
with the pion-nucleon interaction, and $\overline{l}_4$ associated
with $f_\pi$.  This is in addition to the $D_2$ contribution in
the $\siii$ channel.  The allowed regions for $\overline{d}_{18}$ and
$\overline{d}_{16}$ are given in Ref.~\cite{Fettes:fd}, and
$\overline{l}_4$ is known.  Fig.~\ref{fig:1s0mq} shows the presently
allowed values of the scattering length in the $\siii$ channel where
we again have used NDA to estimate the possible values for $D_2$.  It
is clear that for the range of parameters considered the deuteron
could be bound or unbound in the chiral limit, and at present one
cannot make a more definitive statement. 
This last statement is in disagreement with the conclusion of Ref.~\cite{EMmq};
a discussion of this disagreement is given in Ref.~\cite{BSmq}.
The quark-mass dependence of the deuteron binding energy is shown in Fig.~\ref{fig:deutbe}. 
Notice that the $\siii$ scattering length relaxes to
natural values of $\sim 1~{\rm fm}$ as the pion mass is increased beyond 
$\sim 200~{\rm MeV}$ (see Fig.~\ref{fig:1s0mq}). One anticipates similar
behavior in the partially-quenched theory.
Given current uncertainties in
strong interaction parameters, particularly $D_2$, it is at present
unclear whether the same is true in the $\si$ channel (see Fig.~\ref{fig:1s0mq}). 

\begin{figure}[!htb]
\begin{center}
\includegraphics*[width=0.4\linewidth,clip=true]{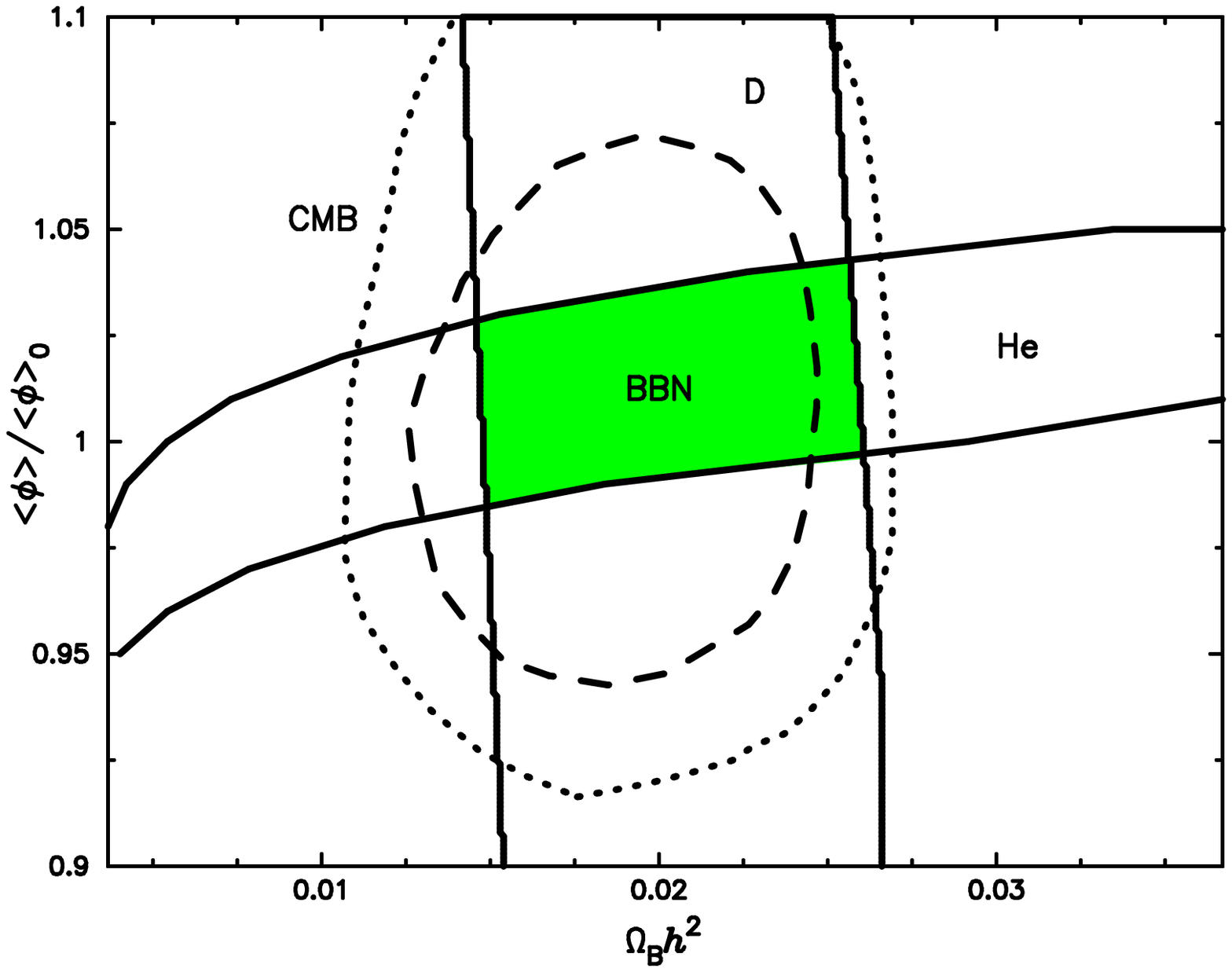}
\hskip 0.6in
\includegraphics*[width=0.4\linewidth,clip=true]{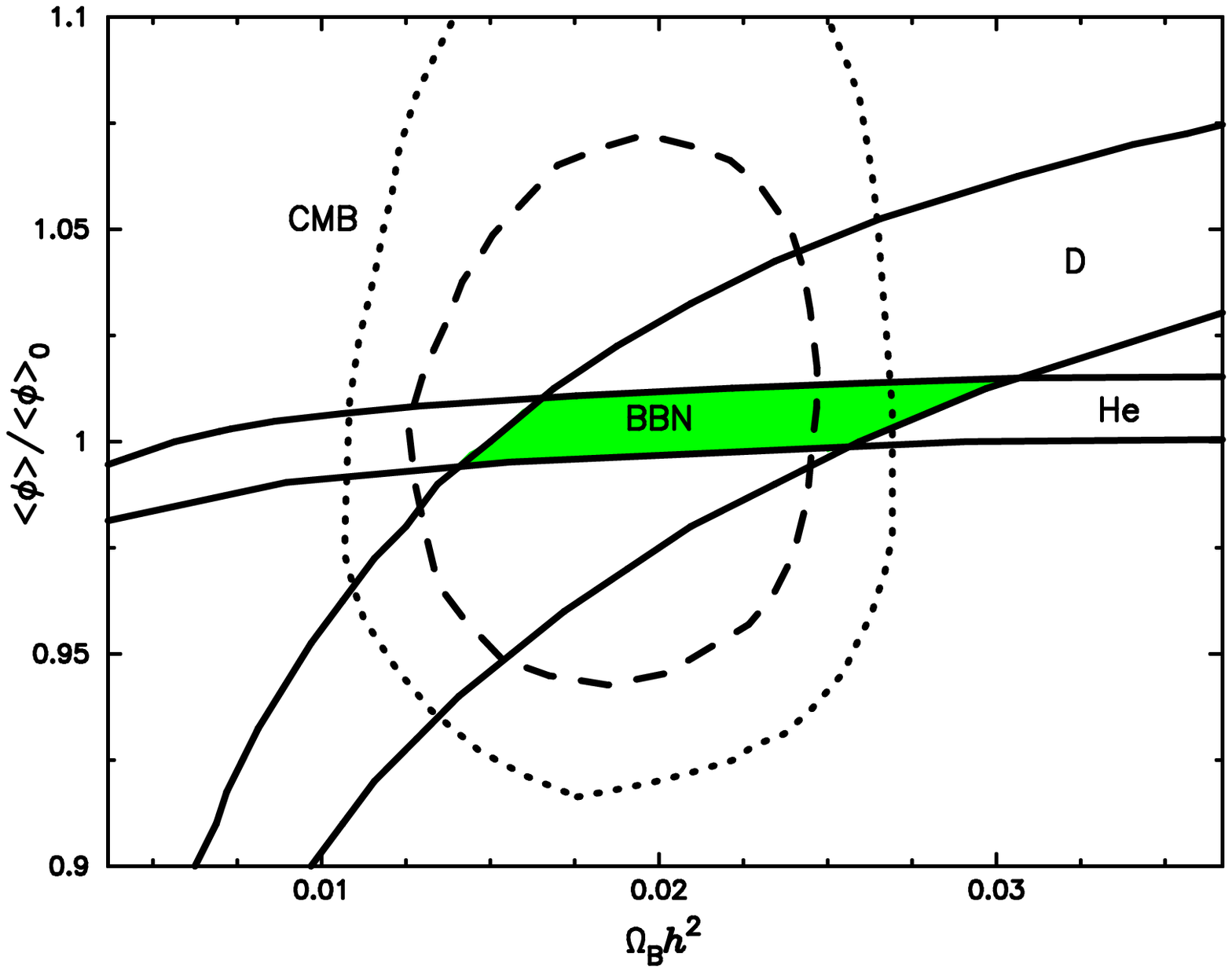}
\end{center}
\vskip -0.5in
\caption{
The solid curves represent the constraints from the $^4$He and
deuterium abundances, assuming no change in the deuteron binding
energy (left panel) and a change in the deuteron binding
energy as shown in Fig.~\ref{fig:deutbe}.
The region allowed by BBN is shaded. }
\label{fig:yoo}
\vskip -0.1in
\end{figure}

As an interesting application of these results, Ref.~\cite{yoo} derive
constraints on the time variation of the Higgs vacuum expectation
value $\left<\phi\right>$ through the effects on BBN, including the
effect of the change in the deuteron binding energy, which alters both
the $^4$He and deuterium abundances significantly. See
Fig.~\ref{fig:yoo}.

It is quite simple to construct the partially-quenched effective field
theory from the known QCD results. And too, it is gratifying to see that one can
obtain analytic results for many observables in the $\si$ channel and in the
higher partial waves. For instance, the $^1 P_1$ scattering volume is given by
\begin{equation}
a (^1 P_1) =  { g_A^2 M_N\over 4\pi f^2 m_\pi^2}
\ +\ 
{g_0^2 M_N\over 12\pi f^2 m_\pi^2}\ {m_{SS}^2-m_\pi^2\over m_\pi^2}\ ,
\end{equation}
where $g_0$ is an axial coupling and $m_{SS}$ is a Goldstone boson
mass containing two sea quarks. Notice that 
the QCD limit agrees with the well-known results~\cite{EWbook}.
One should be concerned about the range of sea and valence quark masses for
which this theory converges.  In QCD it is found that the NN EFT converges for
$m_\pi$ and momenta less than of order $\Lambda_{NN}\sim 300~{\rm MeV}$, and
one suspects that the same radius of convergence will exist in the
partially-quenched theory. If this is indeed the case, lattice calculations
will be required with meson masses of less than $\sim 300~{\rm MeV}$ in order
to match to the EFT and use it to make predictions about nature.  This is
somewhat more restrictive than in the meson and single nucleon sectors and
therefore one would like to see convergent results in those sectors before
being confident in results obtained in the multi-nucleon sectors.

%%%%%%%%%%%%%%%%%%%%%%%%%%%%%%%%%%%%%%%%%%%%%%%%%%%%%%%%%%%%%%%%%%%%
\section{The $\Lambda_Q \Lambda_Q$ Potential}

The lowest-lying baryons containing a single heavy quark can be
classified by the spin of their light degrees of freedom (dof), $s_l$,
in the heavy-quark limit, $m_Q\rightarrow\infty$.  Working with two
light flavors, $u$ and $d$ quarks, the light dof of the isosinglet
baryon, $\Lambda_Q$, have $s_l=0$, while the light dof of the
isotriplet baryons, $\Sigma_Q^{\pm 1,0}$ and $\Sigma_Q^{\pm 1,0 *}$,
(the superscript denotes the third component of isospin) have
$s_l=1$~\footnote{For three light flavors the baryons fall into a
${\bf 6}\oplus \overline{\bf 3}$ of SU(3).}.  In the heavy-quark limit
the spin-${1\over 2}$ $\Sigma_Q^{\pm 1,0}$ baryons are degenerate with
the spin-${3\over 2}$ $\Sigma_Q^{\pm 1,0 *}$ baryons, but are split in
mass from the $\Lambda_Q$ by an amount that does not vanish in the
chiral limit. As the light dof in the
$\Lambda_Q$ have $s_l=0$ in the heavy-quark limit, the light-quark
axial current matrix element vanishes, and thus there is no
$\Lambda_Q\Lambda_Q\pi$ interaction at leading order in the heavy
quark expansion.  This means that there is no OPE (or OEE)
contribution to the $\Lambda_Q\Lambda_Q$ potential in QCD and PQQCD,
and therefore there is no long-distance component in the
$\Lambda_Q\Lambda_Q$ potential.  It is the TPE box and
crossed-box diagrams that provide the longest-distance interaction
between two $\Lambda_Q$'s. In addition, there are local
four-$\Lambda_Q$ operators at the same order in the chiral expansion
but such local interactions give coordinate-space delta-functions.
Analogous diagrams in the two-nucleon sector provide part of the
intermediate-range component of the NN potential.  However, it is
important to realize that there are additional interactions that
contribute to the intermediate-range component of the NN potential in
the chiral expansion, for instance contributions from the
Weinberg-Tomazawa term and also from higher-dimension ${\overline
N}N\pi\pi$ vertices.  Therefore, while the $\Lambda_Q\Lambda_Q$
potential provides a window into the nature of the intermediate-range
NN interaction, it certainly does not provide a complete description.

\begin{figure}[!htb]
\begin{center}
\includegraphics*[width=0.65\linewidth,clip=true]{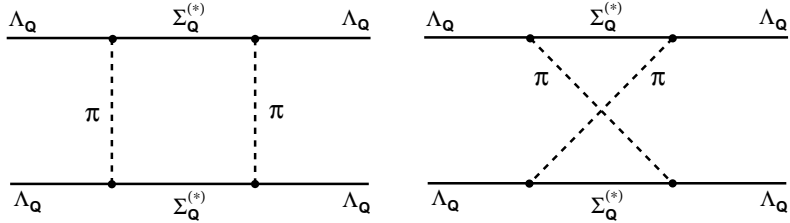}
\end{center}
%\vskip -0.5in
\caption{
The box and crossed-box diagrams that give the longest-distance component of
the $\Lambda_Q\Lambda_Q$ potential. 
Heavy-quark symmetry forbids $\Lambda_Q\Lambda_Q$ intermediate states in the
box and crossed-box diagrams.
 }
\label{fig:BandCB}
\vskip -0.1in
\end{figure}

If the $\Lambda_Q$ and $\Sigma_Q^{(*)}$ were degenerate, we would be
required to solve the coupled-channel system with $\Lambda_Q\Lambda_Q$
and $\Sigma_Q^{(*)}\Sigma_Q^{(*)}$ coupled to $I=0$.  In the charmed
sector the $\Sigma_c -\Lambda_c $ mass splitting is $\Delta=167.1~{\rm
MeV}$ and the $ \Sigma_c^* -\Lambda_c$ mass splitting is
$\Delta=232.7~{\rm MeV}$, and we use the spin-weighted average of
these splittings to estimate $\overline\Delta\sim 211~{\rm MeV}$ in
the heavy-quark limit.  There is no symmetry reason for this
mass-splitting to vanish in the chiral limit, and hence there is no
infrared divergence that requires a coupled-channel analysis.  In the
power-counting we treat $\overline\Delta\sim m_\pi$ and take the
$M_{\Lambda_Q}, M_{\Sigma_Q}\rightarrow \infty$ limit in evaluating
the diagrams in Fig.~\ref{fig:BandCB}.  With this power-counting, one
can directly use the Feynman rules of Heavy-Baryon Chiral Perturbation
Theory (HB$\chi$PT) to describe the low-momentum dynamics of the
nucleon and $\Delta$-resonance, 
without the need to resum the baryon kinetic
energy term as is the case for the box and crossed-box diagrams in the
nucleon sector.  Evaluating the diagrams in Fig.~\ref{fig:BandCB} and
then Fourier transforming them to position space is straightforward~\cite{ABSl}.
For asymptotically large distances, the potential is well-approximated by
\begin{eqnarray}
V^{\rm QCD}(r) & \rightarrow & 
- {3 \ g_3^4\  m_\pi^{9/ 2}\over 16\ \pi^{5/ 2}\  r^{5/ 2}\ \overline\Delta^2 f_\pi^4}
e^{-2 m_\pi r}
\ +\ ...
\ \ \ ,
\end{eqnarray}
which exhibits the expected fall off with a length scale set by
twice the mass of the pion and where the dots represent subleading 
contributions in the large-distance expansion.

\begin{figure}[!htb]
\begin{center}
\includegraphics*[angle=-90,width=0.4\linewidth,clip=true]{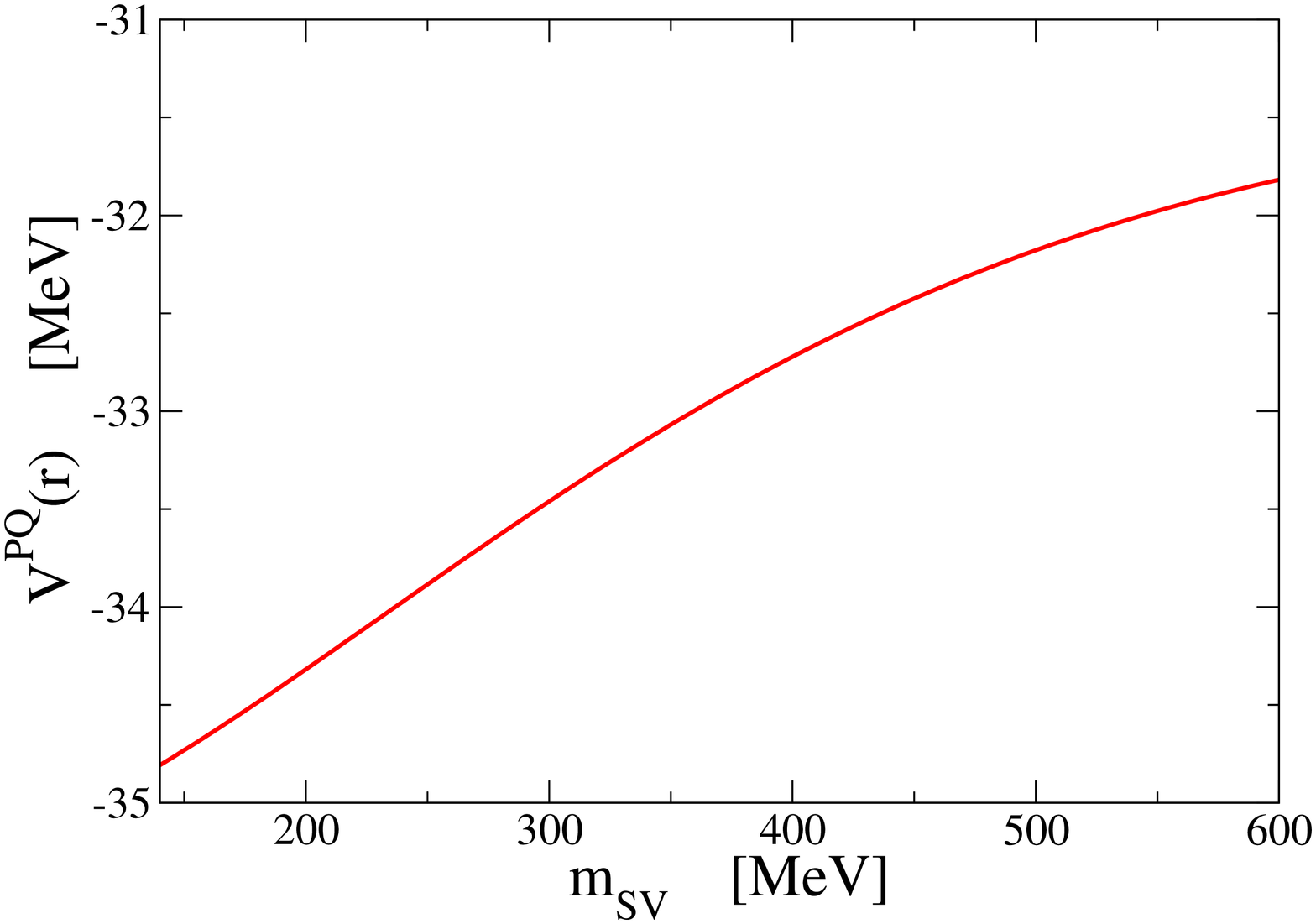}
\hskip 0.6in
\includegraphics*[angle=-90,width=0.4\linewidth,clip=true]{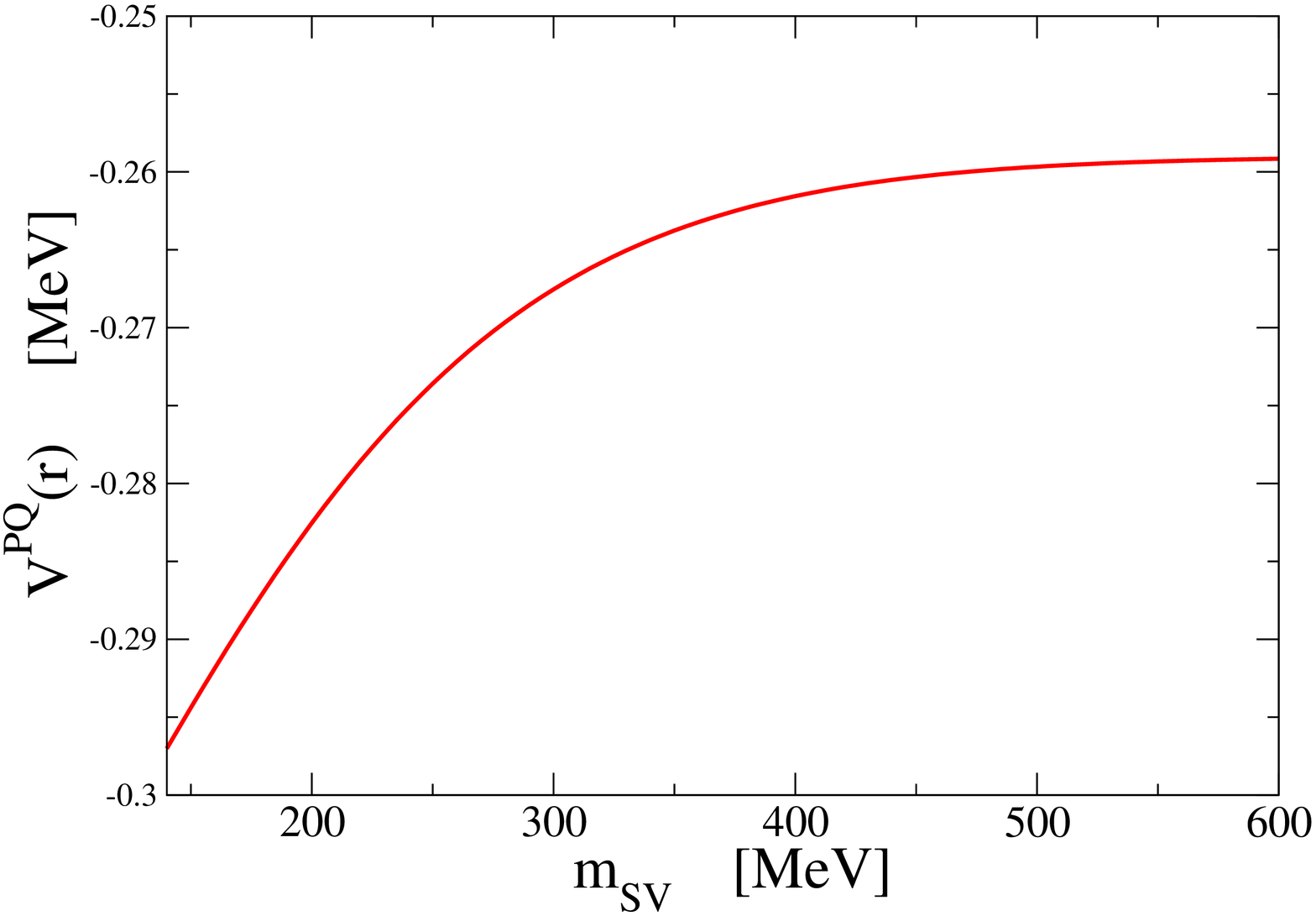}
\end{center}
\vskip -0.5in
\caption{
The left panel shows $V^{\rm PQ}(r)$ evaluated at $r=1~{\rm fm}$ 
as a function of the meson mass $m_{SV}$, while the right panels shows 
$V^{\rm PQ}(r)$ evaluated at $r=2~{\rm fm}$.
The vertical axis is in units of ${\rm MeV}$.
When $m_{SV}=m_\pi$ the value of $V^{\rm PQ}(r)$ is equal to 
$V^{\rm QCD}(r)$.
 }
\label{fig:blah}
\vskip -0.1in
\end{figure}

\begin{figure}[!htb]
\begin{center}
\includegraphics*[width=0.3\linewidth,clip=true]{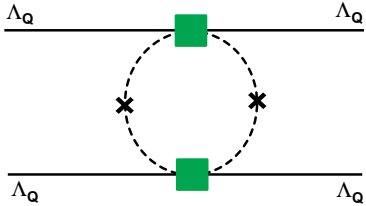}
\end{center}
%\vskip -0.5in
\caption{
The leading hairpin contribution to the $\Lambda_Q\Lambda_Q$ 
potential.}
\label{fig:blah2}
\vskip -0.1in
\end{figure}

The extension of the heavy-baryon sector from QCD to PQQCD is straightforward
and the leading modifications to the $\Lambda_Q\Lambda_Q$ potential are easily
computed~\cite{ABSl}. Fig.~\ref{fig:blah} exhibits the potential as a function
of $m_{SV}$, the mass of the Goldstone boson consisting of a valence quark and
a sea quark, for two values of $r$.
One may wonder whether ``hairpin'' contributions may enter in such
a way as to dominate the potential. The leading hairpin contribution
enters through the diagram in Fig.~\ref{fig:blah2}.
At asymptotically large distances this contribution to the potential becomes
\begin{eqnarray}
V^{\rm HP} (r) & \rightarrow & 
- c_1^2 \ 
{(m_{SS}^2-m_\pi^2)^2  \ m_\pi^{5/2}\over 64\ \pi^{5/ 2}\  
\Lambda_\chi^6} \ 
{e^{-2m_\pi r}\over\sqrt{r}}
\ +\ \ldots
\ \ \ .
\end{eqnarray}
where the dots represent subleading contributions in the large-distance
expansion. While at asymptotically large distances this contribution is larger than that
from the box and crossed-box diagrams, asymptopia finally sets in at distances at which
all contributions are numerically insignificant.

%%%%%%%%%%%%%%
\section{Conclusions}

Understanding how nuclei and nuclear interactions depend upon the light-quark
masses is a fundamental aspect of strong-interaction physics.  Recent work has been
able to explore the $m_q$-dependence of two-nucleon systems using a
recently-developed effective field theory and naive dimensional analysis.  In
the $\si$-channel we expect that di-nucleon systems, such as the di-neutron,
are unbound for all values of $m_q$ less than their physical values.  However,
for $m_q$ larger than their physical values both bound and unbound systems are
presently consistent with data and NDA.  In the $\siii-\diii$ coupled-channels,
where the deuteron resides for the physical values of the quark masses, the
deuteron may or may not be bound in the chiral limit.  A more definitive
statement can only be made with a more precise determination of the $\pi N$
coupling $\overline{d}_{16}$ and a determination of the coefficients of the
leading $m_q$-dependent four-nucleon operators, $D_2$. As discussed in
Ref.~\cite{BSmq}, it is likely that a determination of $D_2$ will
require a future lattice QCD calculation. 

Recent work has computed the potential between two $\Lambda_Q$'s
at leading order in effective field theory in both QCD and
PQQCD. The size of the leading contribution
from hairpin interactions in PQQCD has been estimated. Evidently the
partially-quenched $\Lambda_Q\Lambda_Q$ potential does not suffer from
some of the (partial-) quenching problems that plague the NN potential
due to the absence of single pseudo-Goldstone exchange.  The computed
potentials will allow for the chiral
extrapolation of lattice calculations performed with unphysically
large sea quark masses. As these potentials fall off with a mass scale
set by $\sim 2 m_\pi$, they are quite small for baryon separations
greater than $r\sim 1.5~{\rm fm}$.  Therefore, the theoretical
advantages of studying this system to learn about the NN potential may
be undermined by the difficulties in extracting a signal from lattice
simulations.  However, the simplifications introduced by only having
two light quarks, and a single infinitely-massive quark to fix the
inter-baryon separation makes this system a prime candidate for
studying inter-baryon interactions. It is very exciting
indeed to be so close to making fundamental statements about nuclear physics.

\vskip 0.2in

I would like to thank the organizers for a very enjoyable meeting
and my collaborators Daniel Arndt and Martin Savage for sharing their
insight. This work is supported in part by the U.S. Department of Energy under
Grant No.~DE-FG03-00-ER-41132.

\end{document}